\documentclass[12pt]{revtex4}
\usepackage{graphicx}
\usepackage{epsfig}
\usepackage{amssymb}
\newcounter{fig}
\newcommand{\lbfig}[1]{\refstepcounter{fig} \label{#1} }

\begin{document}


\vskip 0.cm
\rightline{\hskip 3cm ITP-UU-11/38, SPIN-11/29}
\vskip 0.cm

 \title{Symmetry breaking and Goldstone theorem in de Sitter space}

\author{Tomislav Prokopec}
\email{t.prokopec@uu.nl}

\affiliation{Institute for Theoretical Physics, Utrecht University,\\
Postbus 80.195, 3508 TD Utrecht, The Netherlands}

\begin{abstract}

\noindent
We consider an $O(N)$ symmetric scalar field model in the mean field (Hartree)
approximation and show that the symmetry can be broken in de Sitter space.
We find that the phase transition can be of first order, and that its
strength depends non-analytically on the parameters of the model.
We also show that the would-be Goldstone bosons acquire a mass,
effectively becoming pseudo-Goldstone bosons, thus breaking
the $O(N)$ symmetry.
Our results imply that topological defects can form during inflation.

\end{abstract}


\maketitle

 \section{Introduction}
 \label{Introduction}

 Ever since Kirzhnits and Linde~\cite{Kirzhnits:1972ut} pointed out that
thermal radiative effects can induce phase transitions in the early Universe,
phase transitions have played a central role in the early Universe cosmology.
In particular, they have been used to drive out-of-equilibrium
phenomena which can lead to creation of the matter-antimatter asymmetry, preheating,
formation of topological defects, {\it etc.}
The effects induced by particle creation in an expanding Universe setting are
quite delicate and have not yet been fully understood, albeit there is a large
literature on the subject, a non-representative sample includes
Refs.~\cite{Tsamis:1996qm,Tsamis:1996qq,Mukhanov:1996ak,Onemli:2002hr,Onemli:2004mb,Brunier:2004sb,Prokopec:2006ue,Prokopec:2007ak,Prokopec:2008gw,Janssen:2008px,Janssen:2008dw,Prokopec:2002jn,Prokopec:2002uw,Prokopec:2003iu,Kahya:2009sz,Janssen:2009pb}.
Based upon a mean field (Hartree) analysis of a scalar self-interacting theory,
Ford and Vilenkin~\cite{Vilenkin:1982wt,Ford:1985qh}
pointed out a long time ago that
the infrared effects in de Sitter space
may restore symmetries spontaneously broken by the vacuum.
A similar conclusion was reached by Ratra in Ref.~\cite{Ratra:1984yq}.
However, in these works it was not realised that mass generation
can regulate the infrared divergences of de Sitter space.
The infrared effects in (quasi-)de Sitter spaces have received a considerable
attention in recent literature, and several papers have been
published~\cite{Riotto:2008mv,Burgess:2009bs,Burgess:2010dd,Serreau:2011fu,Garbrecht:2011gu}
which have -- just as Ford and Vilenkin --  treated the problem in
the mean field -- or Hartree -- approximation
(see also Refs.~\cite{Reinosa:2011ut,Reinosa:2011cs} for some
recent mean field results on flat space).
The current consensus is that the infrared effects in de Sitter space
are strong enough to restore the broken $O(N)$ symmetry.
While this is correct if the corresponding effective action
is averaged over infinite distances,
from the observational point of view the more relevant
question is whether the symmetry gets broken or restored when averaged
over some fixed physical scale~\footnote{As early as 1986 Ford and
Vilenkin~\cite{Ford:1985qh} correctly reasoned:
"[..] it would be completely consistent with all observations for
our present Universe to be a de Sitter space with $H^{-1}\sim 10^{10}$~yr.
It would be very surprising if this were to have any effect upon symmetry
breaking on terrestrial or subatomic scales." While this observation
is correct, it does not follow from their analysis.}.
In this work we take the point of view that the effective action should
be averaged over some fixed physical scale and we show that symmetries
are then generally not restored in de Sitter space.
We also give a simple criterion for symmetry restoration.
For simplicity, we consider here only the global $O(N)$ symmetric scalar
field model.

 For pedagogical reasons we begin in section~\ref{A real scalar field} by
analysing a real scalar field
($O(1)$ model). The central part of the paper is section~\ref{The O(N) model}
where we analyse an $O(N)$ symmetric model on de Sitter space.
Section~\ref{Discussion} is reserved for a discussion, and
the Appendix for technical details on the de Sitter space
scalar field propagator.

 \section{A real scalar field}
 \label{A real scalar field}

The free action of a real scalar field $\phi(x)$
in $D$ space-time dimensions reads,
\begin{equation}
S[\phi] = \int d^D x \sqrt{-g} \left[-\frac12 g^{\mu\nu} (\partial_\mu\phi)(\partial_\nu \phi)
          -\frac 12 m_0^2 \phi^2 - \frac{\lambda}{4!}\phi^4\right]
\,,
\label{O(1) action}
\end{equation}
where $m_0$ and $\lambda$ denote the field  mass and quartic coupling, respectively,
$g_{\mu\nu}$ is the metric tensor, $g^{\mu\nu}$ its inverse, and
$g={\rm det}[g_{\mu\nu}]$. The metric signature we use is $(-,+,+,..)$.
The vacuum is given by $\phi^2 = 0$ for $m_0^2>0$ and $\phi^2=- 6m_0^2/\lambda$ when
$m_0^2<0$. In the case when $\phi^2>0$ the $Z_2$ symmetry
($\phi\rightarrow -\phi$) of the action~(\ref{O(1) action})
is (completely) broken by the vacuum in which $\phi^2=- 6m_0^2/\lambda$.
 If one goes through a quench from high temperatures ({\it e.g}
 in an early Universe setting),
when $m_0^2<0$ domain walls form by the Kibble mechanism~\cite{Kibble:1980mv}, such that
the state spontaneously breaks translation invariance.
By causality, at least of the order of one domain wall forms per Hubble volume. Once formed,
their energy density scales as $\propto 1/a^2$ ($a$ denotes the scale factor), such
that in decelerating spacetimes domain walls will dominate the energy density at late times
(a menace to get rid of), while in accelerating spacetimes (such as inflation)
they get diluted.

 Here we shall perform a mean field (Hartree) analysis, for which the effective potential
(up to two loop order) is of the form,
\begin{equation}
V_{\rm MF} = V_0 + \frac{1}{2}m_0^2(\phi^2 + \imath \Delta(x;x))
      +\frac{\lambda}{4!}\left(\phi^4 + 6\phi^2\imath \Delta(x;x) +
       3[\imath \Delta(x;x)]^2\right) +\frac{\imath}{2}{\rm Tr}\ln[\imath \Delta(x;x)]
\,,
\label{Veff:MF:1}
\end{equation}
where $m_0^2$ denotes a bare mass term, $\lambda>0$ a quartic coupling,
$\phi(x)=\langle\Omega|\hat \phi(x)|\Omega\rangle$ is a mean field,
$|\Omega\rangle$ is a state,
and $\imath\Delta(x;x^\prime)
= \langle\Omega|T[\delta \hat \phi(x^\prime)\delta\hat \phi(x)]|\Omega\rangle$
is the Feynman propagator for the field fluctuations
$\delta\hat\phi(x)=\hat \phi(x)-\phi(x)$, where $T$ stands for time ordering.
For simplicity, we have assumed that gravity is nondynamical
and that all quantities $\phi(x)$ and $\imath \Delta(x;x)$
are either constant or adiabatically varying in time (on a de Sitter background).
Varying the mean field action
\begin{equation}
 S_{\rm MF}[\phi,\Delta] = \int d^D x \sqrt{-g(x)}
  \left[-\frac 12(\partial_\mu \phi)(\partial_\nu\phi)g^{\mu\nu}
   + \frac 12 [\Box_x \imath \Delta(x;x)]_{x^\prime\rightarrow x}
 - V_{\rm MF}(\phi,\Delta)\right]
\label{MF action}
\end{equation}
with respect to $\phi(x)$ and $\imath \Delta(y;x)$ results in
\begin{eqnarray}
 \left[\Box_x - m_0^2-\frac{\lambda}{2}\imath\Delta(x;x)\right]\phi
   - \frac{\lambda}{6}\phi^3 &=& 0
\label{eom:phi:1}
\\
 \sqrt{-g} \left[\Box_x - m_0^2-\frac{\lambda}{2}(\phi^2+\imath\Delta(x;x))\right]
             \delta^D(x-y) &=& \imath [\imath\Delta(x;y)]^{-1}
\,,
\label{eom:Delta:1}
\end{eqnarray}
where $\Box_x=(-g)^{-1/2}\partial_\mu g^{\mu\nu}\sqrt{-g}\partial_\nu$ denotes
the d'Alembertian ($\Box_x=g^{\mu\nu}\nabla_\mu\nabla_\nu$) as it acts on
a scalar quantity. Since the coincident propagator $\imath \Delta(x;x)$
is in general divergent (see the Appendix),
these equations need to be renormalised. From the structure of
equations~(\ref{eom:phi:1}--\ref{eom:Delta:1}),
a mass renormalisation suffices~\footnote{To renormalise
the mean field effective potential~(\ref{Veff:MF:1}), one also needs to renormalise
$V_0$. Since $V_0$ does not affect our analysis, we shall not
renormalise $V_0$ here.}.
When viewed as a function of the ultraviolet cutoff $\Lambda$,
the coincident propagator exhibits power law divergences $\propto\Lambda^{D-2},\Lambda^{D-4}$,
{\it etc.}~\cite{Reinosa:2011ut,Reinosa:2011cs},
which are automatically subtracted in dimensional regularisation.
In the special cases when the spacetime dimension is even ($D=2n$, $n=1,2,\dots$),
there is a logarithmic divergence in $\Lambda$, that manifests itself
as a simple pole in the coincident propagator,
 $\imath\Delta(x;x)_{\rm div}\propto 1/(D-2n)$ ($n=1,2,\dots$), see
Eqs.~(\ref{A:coincident limit}--\ref{A:propagator:time dep:3})
in the Appendix.
When this divergence is absorbed in the bare mass $m^2_0$,
one gets a finite, renormalised mass term,
\begin{eqnarray}
 m^2&=&m_0^2+\frac{\lambda}{2}\imath\Delta(x;x)_{\rm div}
\nonumber\\
\imath\Delta(x;x)_{\rm div}
  &=&\frac{H^{D-2}\Gamma\left(\frac{D\!-\!1}{2}\right)}{4\pi^{(D-1)/2}}
    \Bigg[\psi\bigg(\frac{D}{2}\bigg)\!-\!\psi(D\!-\!1)
          -\psi\bigg(1\!-\!\frac{D}{2}\bigg)-\gamma_E+\frac{1}{D\!-\!1}
    \Bigg]
\,,
\label{renormalised mass}
\end{eqnarray}
where $m^2$ can be either positive or negative.
Hence, the renormalised, manifestly finite, form of Eqs.~(\ref{eom:phi:1}--\ref{eom:Delta:1})
is,
\begin{eqnarray}
 \left[\Box_x - m^2-\frac{\lambda}{2}\imath\Delta(x;x)_{\rm fin}\right]\phi
   - \frac{\lambda}{6}\phi^3 &=& 0
\label{eom:phi:1ren}
\\
 \sqrt{-g} \left[\Box_x - m^2-\frac{\lambda}{2}(\phi^2+\imath\Delta(x;x)_{\rm fin})\right]
             \imath\Delta(x;x^\prime) &=& \imath\delta^D(x-x^\prime)
\,,
\label{eom:Delta:1ren:B}
\end{eqnarray}
where $\imath\Delta(x;x)_{\rm fin} = \imath\Delta(x;x)-\imath\Delta(x;x)_{\rm div}$
is the finite part of the coincident correlator,
{\it cf.} Eqs.~(\ref{A:coincident limit:2}--\ref{A:propagator:time dep:3}).

 Since $\phi$ is (by assumption) slowly varying, we have $\Box\phi\approx 0$,
and Eq.~(\ref{eom:Delta:1ren:B}) yields
\begin{equation}
\left[m^2+\frac{\lambda}{2}\imath\Delta(x;x)_{\rm fin}+\frac{\lambda}{6}\phi^2\right]\phi = 0
\,.
\label{stationary point:1}
\end{equation}
This is solved by $\phi=0$ or, when $\phi^2>0$, by
\begin{equation}
\phi^2=-\frac{6 m^2}{\lambda}-3\imath\Delta(x;x)_{\rm fin} > 0
\,,
\label{stationary point:2}
\end{equation}
At early times $\imath\Delta(x;x)_{\rm fin}$ grows as given
in~(\ref{A:propagator:time dep}--\ref{A:propagator:time dep:3}), reaching at late
times the de Sitter invariant limit~(\ref{A:coincident limit:2}), and
the condition~(\ref{stationary point:1}) becomes,
\begin{equation}
\phi^2=-\frac{6 m^2}{\lambda}
  -\frac{3\Gamma\left(\frac{D+1}{2}\right)}{2\pi^{(D+1)/2}}\frac{H^D}{m_{\rm MF}^2}> 0
\,,
\label{stationary point:3}
\end{equation}
where $m_{\rm MF}^2=\partial^2 V_{\rm MF}/\partial\phi^2$ is the mean field mass satisfying
the mass gap equation
\begin{eqnarray}
 m_{\rm MF}^2 = m^2 +\frac{\lambda}{2}\Big(\phi^2 + \imath\Delta(x;x)_{\rm fin}\Big)
         =
\cases{ -2m^2 - \lambda \imath\Delta(x;x)_{\rm fin}, & \mbox{if } $\phi^2>0$ \cr
                         \quad\;  m^2 + \frac{\lambda}{2} \imath\Delta(x;x)_{\rm fin},
                                                     & \mbox{if } $\phi^2=0$.\cr
}
\label{mass gap equation}
\end{eqnarray}
Note that the mean field mass $m_{\rm MF}$ is also the mass of field fluctuations
in the correlator equation~(\ref{eom:Delta:1ren:B}).
When Eq.~(\ref{A:coincident limit:2}) is inserted into~(\ref{mass gap equation})
one gets,
\begin{equation}
 m_{\rm MF}^2 +2m^2
+ \frac{\lambda\Gamma\left(\frac{D+1}{2}\right)}{2\pi^{(D+1)/2}}\frac{H^D}{m_{\rm MF}^2} =0
\qquad (\phi^2>0)
\,.
\label{mass gap equation:2}
\end{equation}
This is solved by,
\begin{equation}
 m_{\rm MF\pm}^2 =-m^2 \pm\sqrt{m^4-m_{\rm cr}^4}
 \,,\qquad m_{\rm cr}^4 = (\lambda m_{\rm MF}^2)\imath\Delta(x;x)_{\rm fin}
           = \frac{\lambda H^D\Gamma\left(\frac{D+1}{2}\right)}{2\pi^{(D+1)/2}}
\qquad (\phi^2>0)
\,.
\label{mass gap equation:3}
\end{equation}
We see that -- when the symmetry is broken, $\phi^2>0$ -- there is a minimum $|m^2|$
for which the gap equation~(\ref{mass gap equation:3}) permits a meaningful (real) solution
\footnote{A simple algebra shows that both mean field masses-squared $m_{\rm MF\pm}^2$
in~(\ref{mass gap equation:3}) are positive, and moreover both are consistent
with broken symmetry $\phi^2>0$, see~(\ref{stationary point:3}), which is to
be contrasted to the conclusion in~\cite{Serreau:2011fu}.}:
\begin{equation}
 |m^2| > m_{\rm cr}^2 =
\sqrt{\frac{\lambda H^D\Gamma\left(\frac{D+1}{2}\right)}{2\pi^{(D+1)/2}}}
\,.
\label{critical mass}
\end{equation}
In $D=2,3$ and $4$, $m_{\rm cr}^2 =\sqrt{\lambda/(4\pi)} H,
\sqrt{\lambda/2}H^{3/2}/\pi$, and $\sqrt{3\lambda/2}H^2/(2\pi)$, respectively.
There is a simple way to determine the physically correct sign in
Eq.~(\ref{mass gap equation:3}). In the limit when $H\rightarrow 0$
the infrared enhanced fluctuations are absent,
such that one should recover the tree level mass,
$m_{\rm MF}^2 \rightarrow -2m^2$. This then implies that the physical mean field mass
corresponds to the positive branch in~(\ref{mass gap equation:3}),
\begin{equation}
 m_{\rm MF}^2 =-m^2 + \sqrt{m^4 - m_{\rm cr}^4}
\qquad (\phi^2>0)
\,.
\label{mass gap equation:4}
\end{equation}

\medskip

 On the other hand, when the symmetry is  unbroken, $\phi^2=0$,
Eq.~(\ref{mass gap equation}) implies,
\begin{equation}
 m_{\rm MF}^4 - m^2m_{\rm MF}^2 -\frac12 m_{\rm cr}^4 = 0
\label{mass gap equation:phi2=0}
\end{equation}
which is solved by,
\begin{equation}
 m_{\rm MF}^2 = \frac{m^2}{2} + \sqrt{\frac{m^4}{4}+\frac{m_{\rm cr}^4}{2}}
\qquad (\phi^2=0)
\,.
\label{mass gap equation:phi2=0:2}
\end{equation}
This formula agrees with Eq.~(5.9) of Ref.~\cite{Vilenkin:1982wt}
(provided one makes the replacement $-2\lambda\rightarrow \lambda$
in~\cite{Vilenkin:1982wt}),
and also with the results of Refs.~\cite{Serreau:2011fu,Garbrecht:2011gu}.
In Eq.~(\ref{mass gap equation:phi2=0:2}) we have dropped the solution
with a negative sign in front of the square root, because for that solution
$m_{\rm MF}^2<0$, which is unacceptable on physical grounds
(in this case the de Sitter invariant state would be unstable under
small perturbations).
When the solution~(\ref{mass gap equation:phi2=0:2})
is inserted into Eq.~(\ref{mass gap equation:2}), one can show that
the only real solution for $\phi$ is $\phi=0$, consistent with the
assumption of unbroken symmetry made in deriving
Eq.~(\ref{mass gap equation:phi2=0:2}).

\medskip

 To summarise, we have found that, when
\begin{equation}
m^2<-m_{\rm cr}^2 = -\sqrt{\frac{\lambda H^D\Gamma\left(\frac{D+1}{2}\right)}{2\pi^{(D+1)/2}}}
\qquad ({\rm vacuum\;\; breaks\;\;the \;\; Z_2\;\; symmetry})
\,,
\label{critical mass HA}
\end{equation}
the infrared fluctuations on de Sitter space
may not be able to restore the broken $Z_2$ symmetry ($\phi\rightarrow -\phi$)
of the vacuum of a real scalar field~(\ref{O(1) action}). In this case
$\phi^2=3m_{\rm MF}^2/\lambda>0$ and Eq.~(\ref{mass gap equation:4}) applies.
Otherwise, when $m^2 \geq -m_{\rm cr}^2$ the $Z_2$ symmetry is restored and
Eq.~(\ref{mass gap equation:phi2=0:2}) applies. From Eq.~(\ref{mass gap equation:4})
we see that, in the broken symmetry case, there is a minimum mean field mass, given by
$(m_{\rm MF}^2)_{\rm cr} = -m^2=m_{\rm cr}^2$, implying that for any
finite coupling $\lambda$, as $|m^2|$ increases from $|m^2|<m_{\rm cr}^2$
to $|m^2|>m_{\rm cr}^2$, the order parameter $\phi^2=3m_{\rm MF}^2/\lambda$
will experience a jump,
\begin{equation}
\Delta\phi^2=\frac{3m_{\rm cr}^2}{\lambda}
  = \sqrt{\frac{9H^D\Gamma\left(\frac{D\!+\!1}{2}\right)}{2\lambda\pi^{(D+1)/2}}}
\label{Delta phi:first order transition}
\,.
\end{equation}
Notice that
increasing $H$ and decreasing $\lambda$ strengthens the transition.

 An interesting question is what the above analysis implies for
the history of the Universe, and in particular for inflationary
cosmology. In order to address that question, we shall consider
two scenarios. In {\it Scenario A} inflation starts from a vacuum state with
a non-zero vacuum energy, and $H(t)$ adiabatically decreases in time.
In {\it Scenario B} inflation starts
after an early radiation era,
with a temperature $T\gg H$ ($T\propto 1/a$),
and $H\approx {\rm const.}$ during inflation.

 In {\it Scenario A}, early in inflation the expansion rate $H$ is large
and the criterion~(\ref{critical mass HA})
 is {\it not} met, and hence the symmetry is
unbroken. As $H(t)$ decreases, at some moment $t_{\rm T}$
during inflation $|m^2|=m_{\rm cr}^2(t_{\rm T})$, for $t>t_{\rm T}$
the criterion~(\ref{critical mass HA}) is met, and the symmetry
gets broken. We shall now argue that in this case the transition is
not first order. Based on our de Sitter invariant analysis, one would expect
that at the transition the field acquires the expectation value given
by~(\ref{Delta phi:first order transition}), but the details of the dynamics
of the phase transition are unclear, especially since this has to evoke
a de Sitter breaking physics, which we have so far not discussed.
Nevertheless, based on the existing
literature~\cite{Vilenkin:1983xq,Linde:1991sk,Basu:1991ig,Basu:1992ue}
one can make a crude quantitative analysis. Close to the transition,
the condition for stable defect formation
$m_{\rm MF}^2> 8H^2$~\cite{Basu:1991ig,Basu:1992ue}
is not met. Moreover, since the semiclassical action
is not much greater than one, semiclassical methods do not apply,
and one should make use of Starobinsky's stochastic inflation.
According to Vilenkin~\cite{Vilenkin:1983xq}
and Linde~\cite{Linde:1991sk}, the probability
for nucleation of nearly homogeneous regions (over the horizon size)
of a horizon volume is of the order of one per horizon time and
volume~\footnote{The argument goes as follows. In stochastic inflation,
the probability that the field has a value that is homogeneous over
a Hubble volume is, $P\sim \exp[-\phi^2/(2\langle\delta\phi^2\rangle)]$.
During the transition the value of $\langle\delta\phi^2\rangle$
will grow according to~(\ref{A:propagator:time dep:2})
until it reaches its de Sitter invariant value,
$\langle\delta\phi^2\rangle=3H^4/(8\pi^2m_{\rm MF}^2)$.
A typical value of the field condensate that is unsuppressed will then be
$\phi^2\sim 3H^4/(4\pi^2m_{\rm MF}^2)$, which is $2/3$-rds of
$\Delta \phi^2$ in~(\ref{Delta phi:first order transition})
at the transition.}. This then means that the transition will not be
first order but some higher order or
crossover~\footnote{If initial conditions are $O(N)$ symmetric,
the phase transition will be of some higher (than first) order. If initial
conditions are such to respect $O(N-1)$ (but not $O(N)$ symmetry),
then the transition will be a crossover.}.
Once they nucleate, these regions will grow in size superluminally, and hence
cannot decay; very quickly the whole Universe will be filled by
nearly homogeneous super-Hubble size regions of broken phase.
The field $\phi$ in the Starobinsky's stochastic picture
exhibits a Brownian motion.
The field exhibits random jumps $\delta \phi\sim H/(2\pi)$
over time scales $\sim 1/H$, such that it will achieve
the broken phase value $\phi\sim H/\lambda^{1/4}$
after ${\cal O}(1/\sqrt{\lambda})$ jumps.
This conclusion is supported by a more quantitative analysis which
follows from Eq.~(\ref{A:propagator:time dep:2}),
according to which (in $D=4$)
$m^2_{\rm MF}\simeq (\lambda/2)[H^2/(4\pi^2)]\ln(a)$
and will reach $m_{\rm cr}^2=\sqrt{(3\lambda/2)}H^2/(2\pi)$
after $\ln[a(t)/a(t_{\rm T})]\simeq 4\pi\sqrt{3/(2\lambda)}$
$e$-foldings (assuming it started from zero),
at which point the field will settle to its
critical value~(\ref{Delta phi:first order transition}).
Of course, as $H$ adiabatically evolves, $\phi$ will adjust
to $\phi^2=3m_{\rm MF}^2/\lambda$, with  $m_{\rm MF}^2$
given in~(\ref{mass gap equation:4}).

 The above analysis shows that there is effectively no barrier
to growth of $\phi$, and it will grow as required by the local dynamics.
But, there is still the concern that causality will prevent growth of
a condensate of constant value accross causally disconnected regions
of de Sitter space.
Indeed, the size of domains of constant $\phi$ will at any given time
in inflation be finite, and averaging over the whole Universe will
necessarily give $\phi=0$. The physical size of domains at time $t>t_T$
will be $\sim {\rm e}^{H(t-t_{\rm T})}/H$, and as long as this size is larger
than any physical size of relevance, one can take
$\phi$ to be constant (spatially independent) accross the whole Universe.
But, what about the correlator, which is formally obtained by
averaging field fluctuations over the whole Universe?
For a massive scalar, a typical physical scale over which field fluctuations
significantly contribute is given by $m_{\rm MF}^{-1}$ or smaller,
and contribution of larger scales is suppressed. Hence, as long as
the scale $(1/H)\exp[H(t-t_{\rm T})]$ is much larger than $1/m_{\rm MF}$,
there is a well defined separation of scales in between
the condensate $\phi$ and the fluctuations
 $\imath\Delta(x;x)=\langle\delta\phi^2\rangle$,
and the averaging procedure advocated in this paper is justified.
Moreover, when the field is massive, one can extend the infrared
cutoff $\sim m_{\rm MF}^{-1}$
(which was there due to the finite size of averaging domain)
to zero without significantly changing the result for
$\langle\delta\phi^2\rangle$, and with the bonus of restoring
de Sitter invariance.

 In {\it Scenario B} inflation is preceded by a radiation
era characterised by a temperature $T\gg H$, such that,
at early stages of inflation, $m_{\rm MF}^2\simeq m^2 +\lambda T^2/24>0$
and the symmetry is unbroken. During inflation
the temperature drops rapidly as $T\propto 1/a\propto {\rm e}^{-Ht}$,
and after some time $m^2 +\lambda T^2/24<0$.
This scenario is a realisation of quenched transition mentioned
at the beginning of this section.
If at that moment the criterion for domain wall formation
$|m^2|>4H^2$ is met, the semiclassical analysis of
phase transition applies~\footnote{The semiclassical treatment
predics the following probability for nucleation of Hubble-size
domains,
$P\sim \exp\Big(-\frac{2\pi^2m_{\rm MF}^4}{3\lambda H^4}\Big)\simeq
 \exp\Big(-\frac{8\pi^2 |m^2|^2}{3\lambda H^4}\Big)\ll 1$~\cite{Linde:1991sk},
where to get the last equality we used $m_{\rm MF}^2\simeq 2|m^2|$.
Since $|m^2|>4H^2$, the expression in the exponent is much greater
than one, signaling applicability of the semiclassical approximation.},
and the transition will be of {\it first order\/}:
bubbles of broken phase will nucleate
and moreover topological defects will form.
A detailed description of production and evolution of spherical domain walls,
loops of (global) cosmic strings and global monopole-antimonopole pairs
during inflation is given in Refs.~\cite{Basu:1991ig,Basu:1993rf}.

In summary, we have considered two inflationary scenarios. In
{\it Scenario A} the transition is a crossover,
a temporary breakdown of de Sitter symmetry occurs,
and a (approximate) de Sitter invariant state is reached after some time
after the transition.
In {\it Scenario B} inflation is preceded by a radiation era,
a first order transition can occur and
topological defects such as domain walls may form.

 \section{The $O(N)$ model}
 \label{The O(N) model}

We shall now consider the symmetry breaking in an $O(N)$ symmetric scalar field theory
in the early Universe setting. Recall that this model allows for formation
of (global) cosmic strings (when $N=2$), global monopoles (when $N=3$),
global cosmic textures (when $N=4$), {\it etc}.
The free action of an $O(N)$ symmetric scalar field reads,
\begin{equation}
S = \int d^D x \sqrt{-g} \left[-\frac12 g^{\mu\nu}\sum_{a=1}^N (\partial_\mu\phi_a)(\partial_\nu \phi_a)
    -\frac 12 m_0^2 \sum_{a=1}^N\phi_a^2 - \frac{\lambda}{4N}\Bigg[\sum_{a=1}^N\phi_a^2\Bigg]^2
      \right]
\,.
\label{O(N) action}
\end{equation}
Similarly as in the Brout-Englert-Higgs (BEH) mechanism,
when $m_0^2<0$ and $\lambda>0$ the vacuum breaks the $O(N)$ symmetry to
the $O(N-1)$ symmetry, such that the resulting vacuum manifold is
the $N$ dimensional sphere, $S^N \sim O(N)/O(N\!-\!1)$~\cite{Kibble:1980mv}.
As a result, one of the scalars acquires a mass, while the other $N\!-\!1$ scalars
remain massless. These massless excitations are known as Goldstone bosons.
Unlike in the simple $O(N)$ model~(\ref{O(N) action}), at low energies
the Goldstone bosons in the BEH mechanism acquire a mass and
become the longitudinal excitations of the $W^\pm$ and $Z$ bosons.
For that reason they are known as pseudo-Goldstone bosons. We shall now see
that the Goldstone bosons of the $O(N)$ model in de Sitter space (more generally
in inflationary spacetimes) become massive due to the infrared (super-Hubble) enhancement
of scalar correlations, and in that respect they can be considered as
pseudo-Goldstone bosons.

 The mean field (two loop) effective potential of an $O(N)$ symmetric field $\phi_a$
corresponding to the tree level action~(\ref{O(N) action}) reads,
\begin{eqnarray}
V_{\rm MF} &=&
    \frac 12 m_0^2 \left[\sum_{a=1}^N\left(\phi_a^2 +\imath\Delta_{aa}(x;x)\right)\right]
    + \frac{\lambda}{4N}
    \Bigg[
       \Bigg(\sum_{a=1}^N\phi_a^2\Bigg)^2
       +2\Bigg(\sum_{a=1}^N\phi_a^2\Bigg)\sum_{b=1}^N\imath\Delta_{bb}(x;x)
\label{O(N) effective potential}\\
       &+&4\sum_{a,b=1}^N\phi_a\phi_b\imath\Delta_{ab}(x;x)
       +\Bigg(\sum_{a=1}^N\imath\Delta_{aa}(x;x)\Bigg)^2
       +2\sum_{a,b=1}^N\bigg(\imath\Delta_{ab}(x;x)\bigg)^2
    \Bigg]
    +\frac{\imath}{2}{\rm Tr}\ln\Big(\imath\Delta_{aa}(x;x)\Big)
\,,
\nonumber
\end{eqnarray}
resulting in the following two loop effective action,
\begin{equation}
S_{\rm MF}[\phi_a,\Delta_{bc}] = \int d^D x \sqrt{-g}
   \left[-\frac12 \sum_{a=1}^N g^{\mu\nu}(\partial_\mu\phi_a)(\partial_\nu \phi_a)
       + \frac{1}{2}\sum_{a=1}^N[\Box_x\imath\Delta_{aa}(x;x^\prime)]_{x^\prime\rightarrow x}
       - V_{\rm MF}    
   \right]
\,.
\label{O(N) action:MF}
\end{equation}
Varying the action with respect to $\phi_a(x)$ and $\Delta_{ba}(x^\prime;x)$
gives the following (mean field) equations of motion
({\it cf.} Eqs.~(\ref{eom:phi:1ren}--\ref{eom:Delta:1ren:B})),
\begin{eqnarray}
  \left[\Box_x - m_0^2-\frac{\lambda}{N}\sum_{b=1}^N
              \left(\phi_b^2+\imath\Delta_{bb}(x;x)\right)\right]\phi_a(x)
          - \frac{2\lambda}{N}\sum_{b=1}^N\imath \Delta_{ab}(x;x)\phi_b(x) &=& 0
\label{O(N):eom phi}
\\
   \left[\Box_x - m_0^2
     - \frac{\lambda}{N}\sum_{c=1}^N\left(\phi_c^2+\imath\Delta_{cc}(x;x)\right)
    \right]\imath\Delta_{ab}(x;x^\prime)
  \hskip 3.6cm &&
\nonumber\\
   -\,\frac{2\lambda}{N}\sum_{c=1}^N\left[\phi_a(x)\phi_c(x)
                          +\imath\Delta_{ac}(x;x)\right]\imath\Delta_{cb}(x;x^\prime)
                 &=& \delta_{ab}\frac{\imath\delta^D(x-x^\prime)}{\sqrt{-g}}
\,,
\label{O(N):eom Delta}
\end{eqnarray}
where
\begin{equation}
\imath\Delta_{cb}(x;x^\prime)
     =\left\langle\Omega|T[\delta\hat\phi_b(x^\prime)\delta\hat\phi_a(x)]|\Omega\right\rangle
\,
\nonumber
\end{equation}
denotes the time-ordered scalar field propagator for scalar field fluctuations,
$\delta\hat\phi_a(x)=\hat\phi_a(x)-\langle\Omega|\hat\phi_a(x)|\Omega\rangle$.
Just as in the one scalar case,
Eqs.~(\ref{O(N):eom Delta}) can be renormalised by absorbing the infinite part of the
coincident scalar propagator~(\ref{A:coincident limit:2}) into the bare mass $m_0^2$,
\begin{equation}
m^2=m_0^2+\frac{(N+2)\lambda}{N}\frac{H^{D-2}\Gamma\left(\frac{D\!-\!1}{2}\right)}{4\pi^{(D-1)/2}}
    \left[\psi\left(\frac{D}{2}\right)\!-\!\psi(D\!-\!1)
          \!-\!\psi\left(1\!-\!\frac{D}{2}\right)\!-\!\gamma_E +\frac{1}{D\!-\!1}
    \right]
\,,
\label{mass renormalisation}
\end{equation}
resulting in manifestly finite renormalised equations analogous to
Eqs.~(\ref{eom:phi:1ren}--\ref{eom:Delta:1ren:B}).

 In the limit when the fields are slowly varying ($\Box\phi_a\simeq 0$),
the renormalised form of
 Eq.~(\ref{O(N):eom phi}) yields the following criterion for symmetry breaking,
\begin{equation}
\Bigg[m^2+\frac{\lambda}{N}\sum_{b=1}^N\left(\phi_b^2+\imath\Delta_{bb}(x;x)_{\rm fin}\right)\Bigg]\phi_a(x)
          +\frac{2\lambda}{N}\sum_{b=1}^N\imath \Delta_{ab}(x;x)_{\rm fin}\phi_b(x) = 0
\,.
\label{O(N):symmetry breaking}
\end{equation}
The field mass matrix $M^2$ is obtained by taking a second field derivative of $V_{\rm MF}$
(or equivalently by taking a single derivative with respect to $\imath\Delta_{ba}$),
\begin{equation}
M^2_{ab}  = \Bigg[m^2+\frac{\lambda}{N}\sum_{c=1}^N\left(\phi_c^2+\imath\Delta_{cc}(x;x)_{\rm fin}\right)\Bigg]\delta_{ab}
   +\frac{2\lambda}{N}\left[\phi_a\phi_b+\imath\Delta_{ab}(x;x)_{\rm fin}\right]
\,.
\label{mass matrix}
\end{equation}
Notice that this result can be read off also from Eq.~(\ref{O(N):eom Delta}), as $M^2$ is the mass term of the propagator $\imath\Delta_{ab}$.
Both Eq.~(\ref{O(N):symmetry breaking}) and~(\ref{mass matrix}) contain in general off-diagonal terms.
One can diagonalize them by an $N\times N$ dimensional orthonormal matrix $R=(R_{ab})$, $R\cdot R^T=I$, for which
\begin{equation}
 \phi_a^d = \sum_{b}R_{ab}\phi_b = \phi\left(\begin{array}{c}1 \cr
                                                         0 \cr
                                                         \vdots \cr
                                                         0 \cr
                                         \end{array}\right)
\,,\qquad    \imath\Delta_{ab}^d = \sum_{ce}R_{ac}\imath\Delta_{ce}R_{be} = \left(\begin{array}{ccccc}
                                                         \imath\Delta^d_{11} & 0 & 0 & \cdots & 0\cr
                                                         0 & \imath\Delta^d_{22} & 0 & \cdots & 0\cr
                                                         \vdots & \vdots  & \dots & \vdots & \vdots \cr
                                                         0 & 0 & \dots & 0 & \imath\Delta^d_{NN}\cr
                                         \end{array}\right)
\,.
\nonumber
\end{equation}
Because of the unbroken $O(N-1)$ symmetry, $\imath\Delta^d_{ii}$ are all equal for $2\leq i\leq N$.
The diagonal form of Eqs.~(\ref{O(N):symmetry breaking}--\ref{mass matrix}) is:
\begin{equation}
\Big[m^2+\frac{\lambda}{N}\left(\phi^2+3\imath\Delta^d_{11}(x;x)_{\rm fin}+(N-1)\imath\Delta^d_{22}(x;x)_{\rm fin}\right)\Big]\phi^d_1(x)= 0
\,.
\label{O(N):symmetry breaking:2}
\end{equation}
($\phi_i^d=0$ for $2\leq i\leq N$) and
\begin{eqnarray}
M_1^2\equiv (M^d_{11})^2  &=& m^2+\frac{\lambda}{N}
            \left[3\phi^2+3\imath\Delta^d_{11}(x;x)_{\rm fin}+(N-1)\imath\Delta^d_{22}(x;x)_{\rm fin}\right]
\nonumber\\
M_g^2\equiv(M^d_{ii})^2  &=& m^2+\frac{\lambda}{N}
            \left[\phi^2+\imath\Delta^d_{11}(x;x)_{\rm fin}+(N+1)\imath\Delta^d_{22}(x;x)_{\rm fin}\right]
\,,\qquad (2\leq i\leq N)
\,.
\label{mass matrix:2}
\end{eqnarray}
 Eq.~(\ref{O(N):symmetry breaking:2}) implies that the $O(N)$ symmetry is broken when
\begin{equation}
\phi^2=(\phi_1^d)^2=-\frac{Nm^2}{\lambda}-3\imath\Delta^d_{11}(x;x)_{\rm fin}-(N-1)\imath\Delta^d_{22}(x;x)_{\rm fin}> 0
\,.
\label{O(N):symmetry breaking:3}
\end{equation}
Otherwise, $\phi_1^d=0$ and the symmetry is unbroken. When this is inserted into~(\ref{mass matrix:2}), we get that the mass
terms (in the broken phase) become,
\begin{eqnarray}
M_1^2\equiv (M^d_{11})^2  &=& \frac{2\lambda}{N}\phi^2 = -2m^2-\frac{2\lambda}{N}
            \left[3\imath\Delta^d_{11}(x;x)_{\rm fin}+(N-1)\imath\Delta^d_{22}(x;x)_{\rm fin}\right]
\nonumber\\
M_g^2\equiv(M^d_{ii})^2  &=& \frac{2\lambda}{N}\left[\imath\Delta^d_{ii}(x;x)_{\rm fin}-\imath\Delta^d_{11}(x;x)_{\rm fin}\right]
\,,\qquad (2\leq i\leq N)
\,.
\label{mass matrix:3}
\end{eqnarray}
In the special case when $N=1$ the first equation agrees with Eq.~(\ref{mass gap equation})
(provided, of course, one rescales the $\lambda$ in~(\ref{mass matrix:3}) as
 $\lambda\rightarrow \lambda/6$).

With this, the renormalised and diagonalised form of Eq.~(\ref{O(N):eom Delta}) becomes,
\begin{eqnarray}
   \left[\Box_x - M_1^2\right]\imath\Delta^d_{11}(x;x^\prime) = \frac{\imath\delta^D(x\!-\!x^\prime)}{\sqrt{-g}}
\,,\qquad
   \left[\Box_x - M_g^2\right]\imath\Delta^d_{ii}(x;x^\prime) = \frac{\imath\delta^D(x\!-\!x^\prime)}{\sqrt{-g}}
\quad (2\leq i \leq N)\,.
\quad
\label{O(N):eom Delta:ren:diag}
\end{eqnarray}
The implied stability of de Sitter space then demands
that both $(M_{ii}^d)^2=M_g^2>0$ ($N\geq i\geq 2$)
and $M_1^2>0$. Next we insert the coincident propagator~(\ref{A:coincident limit:2})
into~(\ref{mass matrix:3}) to obtain:
\begin{eqnarray}
M_1^2 &=& -2m^2-\frac{\lambda H^D\Gamma\Big(\frac{D+1}{2}\Big)}{N\pi^{(D+1)/2}}
            \left[\frac{N\!-\!1}{M_g^2} + \frac{3}{M_1^2}\right]
\nonumber\\
M_g^2 &=& \frac{\lambda H^D\Gamma\Big(\frac{D+1}{2}\Big)}{N\pi^{(D+1)/2}}\left[\frac{1}{M_g^2} - \frac{1}{M_1^2}\right]
\,.
\label{mass matrix:4}
\end{eqnarray}
Notice that positivity of $M_g^2$ implies that $M_1^2>M_g^2>0$.
The fact that the Goldstone bosons become massive on de Sitter background is reminiscent
of the BEH mechanism, in which the Goldstones are `eaten up' by the longitudinal
degrees of freedom of the $W^\pm$ and $Z$ bosons, thus becoming massive.
Moreover, massive pseudo-Goldstones imply a complete breaking of the original
$O(N)$ symmetry of the model.

Equations~(\ref{mass matrix:4})
 are the main result of this work. In order to analyse them, it is convenient to
work with the following dimensionless quantities,
\begin{equation}
  \mu_1^2 = \frac{M_1^2}{(-2m^2)}\,,\qquad
    \mu_g^2 = \frac{M_g^2}{(-2m^2)}\,,\qquad
    \lambda_D = \frac{\lambda H^D\Gamma\Big(\frac{D+1}{2}\Big)}{N\pi^{(D+1)/2}(-2m^2)^2}
\,,
\label{rescalings}
\end{equation}
after which Eqs.~(\ref{mass matrix:4}) become
\begin{equation}
\mu_1^2 = 1-\lambda_D\left(\frac{N\!-\!1}{\mu_g^2} + \frac{3}{\mu_1^2}\right)
\,,\qquad
\mu_g^2 = \lambda_D\left(\frac{1}{\mu_g^2} - \frac{1}{\mu_1^2}\right)
\,.
\label{mass matrix:5}
\end{equation}
 Before we perform a general analysis of these equations, notice that in the case when $N=1$,
 the second equation decouples, and one gets
\begin{equation}
 \mu_1^4 -\mu_1^2 + 3\lambda_D = 0
\,,
\nonumber
\end{equation}
whose (physical) root~\footnote{Recall that, in the limit when $\lambda_D\rightarrow 0$,
the physical branch yields $\mu_1^2 = M_1^2/(-2m^2) = 1$.}
is
\begin{equation}
\mu_1^2 = \frac12 +\sqrt{\frac14 - 3\lambda_D}
\,.
\label{mass 1:N=1}
\end{equation}
The minimum critical mass is then determined by $\lambda_D<(\lambda_D)_{\rm cr}=1/12$
($1\geq \mu_1^2\geq (\mu_1^2)_{\rm cr}=1/2$),
which accords with Eq.~(\ref{critical mass}) (when one takes account of
the different definition of $\lambda$ in the Lagrangians~(\ref{O(1) action})
 and~(\ref{O(N) action}))~\footnote{
Formally, one can also solve Eq.~(\ref{mass matrix:5}) for $\mu_g^2$ in the $N=1$ case,
and one finds for the critical value, $(\mu_g^2)_{\rm cr} = (\sqrt{13}-1)/12 \simeq 0.217$,
which agrees with the results plotted in figures~\ref{fig:masses}
and~\ref{fig:masses vs N and Lambda}.}.

 In the general case the gap equations~(\ref{mass matrix:5}) admit a small coupling
expansion. Similarly as in the thermal case, the expansion parameter is $\sqrt{\lambda_D}$,
and hence non-perturbative,
\begin{eqnarray}
  \mu_1^2 &=& 1 -(N\!-\!1)\sqrt{\lambda_D} -\frac{N\!+\!5}{2}\lambda_D
          -\frac{(N\!-\!1)(4N\!-\!21)}{8}\lambda_D^{3/2}+{\cal O}(\lambda_D^2)
\nonumber\\
  \mu_g^2 &=& \sqrt{\lambda_D} -\frac{1}{2}\lambda_D
          -\frac{4N\!-\!5}{8}\lambda_D^{3/2}+{\cal O}(\lambda_D^2)
\,.
\label{mus:series}
\end{eqnarray}
\begin{figure}[htbp]
 \centering
  \epsfig{file=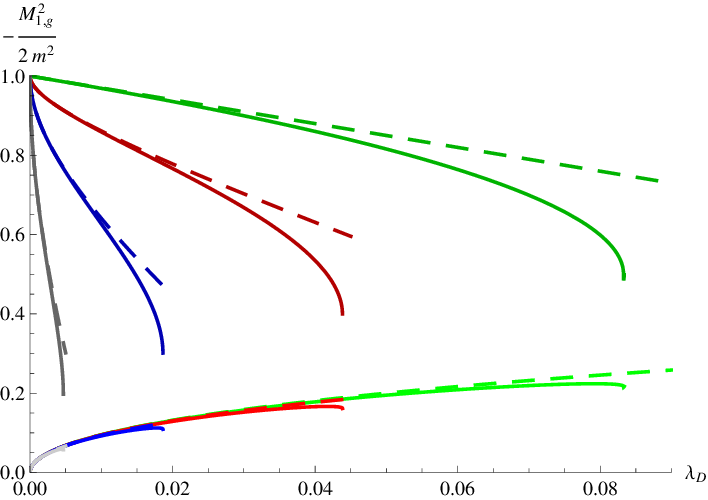, height=4.in,width=5.1in}
   {\small\rm \caption{The rescaled masses $\mu_1^2=M^2_1/(-2m^2)$ (upper curves)
    and $\mu_g^2=M^2_g/(-2m^2)$ (lower curves) as a function of
    the rescaled (dimensionless) coupling $\lambda_D$ defined in~(\ref{rescalings}).
    Both the exact solutions (solid) of Eqs.~(\ref{mass matrix:5})
    and the approximate solutions~(\ref{mus:series}) (dashed) are shown.
    When viewed from the right to left,
    we have plotted the cases: $N=1$ (green),
    $N=2$ (red), $N=4$ (blue) and $N=10$ (gray).
   }}
   \lbfig{fig:masses}
\end{figure}
In figure~\ref{fig:masses}
we show the rescaled masses $\mu_1^2$ and $\mu_g^2$ as a function of $\lambda_D$.
Both, the small coupling series solutions~(\ref{mus:series}) (dashed)
as well as the full solutions of~(\ref{mass matrix:5}) (solid) are shown.
We have plotted the heavy (Higgs) scalar mass $\mu_1^2=M_1^2/(-2m^2)$
(the upper curves starting at one when $\lambda_D=0$)
and the pseudo-Goldstone mass $\mu_g^2=M_g^2/(-2m^2)$
(the lower curves starting at zero when $\lambda_D=0$) for $N=1$
(the most extended green curves),
$N=2$ (the intermediate red curves), $N=4$ (the intermediate blue curves) and $N=10$
(the most squeezed gray curves). While the approximate solutions~(\ref{mus:series})
continue for a while longer, the exact solutions end suddenly at a critical point,
at which a minimum (critical) mass is reached.
Just like for the case when $N=1$, where the mass parameter
decreases monotonically from $\mu_1^2=1$ (at $\lambda_D=0$)
to $(\mu_1^2)_{\rm cr}=1/2$ (at $\lambda_D=(\lambda_D)_{\rm cr}=1/12$),
see Eq.~(\ref{mass 1:N=1}), for $N>1$, $\mu_1^2$ evolves monotonically
from $\mu_1^2=1$ (at $\lambda_D=0$)
to some $(\mu_1^2)_{\rm cr}>0$, the end point being a function of $N$.

\begin{figure}[htbp]
 \centering
  \epsfig{file=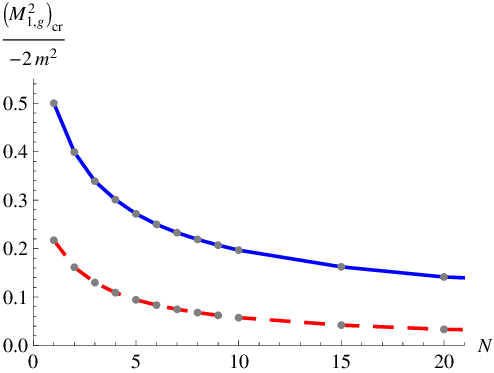, height=2.9in,width=3.4in}
  \epsfig{file=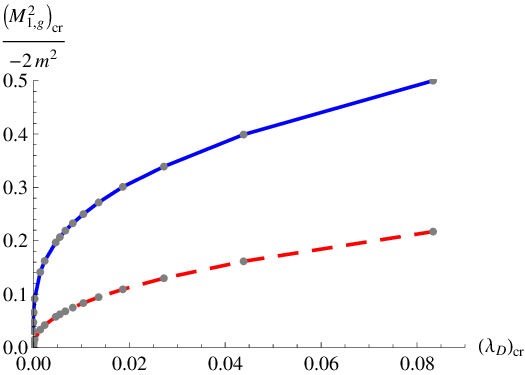, height=2.9in,width=3.4in}
   {\small\rm \caption{The rescaled critical (minimum) masses
     for the Higgs-like excitation $(\mu_1^2)_{\rm cr}=(M_1^2)_{\rm cr}/(-2m^2)$
    (the upper solid blue curve) and for the pseudo-Goldstones
     $(\mu_g^2)_{\rm cr}=(M_g^2)_{\rm cr}/(-2m^2)$ (the lower dashed red curve)
     as a function of $N$ (left panel) and as a function of the rescaled critical coupling
     $(\lambda_D)_{\rm cr}$ (defined as the maximum allowed $\lambda_D$ for a given $N$)
     (right panel). Individual points where $N$ is an integer are shown as gray dots.}}
   \lbfig{fig:masses vs N and Lambda}
\end{figure}
As we have seen in the analysis of the one scalar field case,
this end point plays
an important role as it tells us how the system behaves at
the critical point (where the transition takes place).
In order to study the critical behaviour in some detail,
in figure~\ref{fig:masses vs N and Lambda} we plot
the critical mass parameters $(\mu_1^2)_{\rm cr}$
(upper solid blue curve) and $(\mu_g^2)_{\rm cr}$ (lower dashed red curve)
as a function of $N$ (left panel) and as a function of
$(\lambda_D)_{\rm cr}={\rm max}[\lambda_D]$ for a fixed $N$
(right panel) (corresponding to the end points on figure~\ref{fig:masses}).
One can show that, to a good approximation,
$(\lambda_D)_{\rm cr}\simeq 3/[2(N+2)^2]$
(more precisely $(\lambda_D)_{\rm cr}\simeq 1.62/[N+2]^{1.92}$),
such that $(\lambda_D)_{\rm cr}$ approaches approximately quadratically
zero as $N\rightarrow \infty$. A manifestation of this is the fractional power behaviour of $\mu_{1,g}^2$
close to the origin for small $(\lambda_D)_{\rm cr}$ (large $N$) seen on the
right panel in figure~\ref{fig:masses vs N and Lambda}.

Just as in the one field case,
the transition in the $O(N)$ model is such that, imposing de Sitter symmetry,
implies a jump in the order parameter at the transition.
Indeed, from Eq.~(\ref{O(N):symmetry breaking:2}) we see that
\begin{equation}
 \Delta\phi^2 = \frac{N(M_1^2)_{\rm cr}}{2\lambda}
              = \frac{N(-m^2)(\mu_1^2)_{\rm cr}}{\lambda} > 0
\,,
\label{phase transition:O(N)}
\end{equation}
with $(\mu_1^2)_{\rm cr}$ plotted in figure~\ref{fig:masses vs N and Lambda}.
In fact, from the left and right panels on
figure~\ref{fig:masses vs N and Lambda} one can
read off that $(\mu_1^2)_{\rm cr}\approx 2/[3\sqrt{N}]$ and
$(\mu_1^2)_{\rm cr}\propto (\lambda_D)_{\rm cr}^{1/4}\propto (\lambda/N^2)^{1/4}|m^2|^{-1/2}$,
implying that $\Delta\phi^2\propto |m| N^{1/2}/\lambda^{3/4}$, which is to be compared with
the single field result~(\ref{Delta phi:first order transition}), where we found
that $(\Delta\phi^2)_{N=1}\propto \lambda^{-1/2}$. Hence,
in the large $N$ limit the strength of the transition~(\ref{phase transition:O(N)})
exhibits a qualitatively different dependence on $\lambda$ and $m^2$
than in the single field case~(\ref{Delta phi:first order transition}).
Ref.~\cite{Serreau:2011fu} considered the analogous problem in the large $N$
limit and found that no jump in the order parameter is possible.
The results can be related to ours by noticing that the masses we found
scale as $M_{1,g}^2\propto 1/\sqrt{N}$
({\it cf.} figure~\ref{fig:masses vs N and Lambda})
and thus vanish as $N\rightarrow \infty$, and hence the (rescaled) order
parameter~(\ref{phase transition:O(N)}),
$\Delta\phi^2/N\propto 1/\sqrt{N}$ also vanishes as $N\rightarrow \infty$.

\medskip

 For completeness, we shall now briefly analyse the unbroken symmetry case.
In this case $\phi_i^d=0$ ($i=1,2,..,N$) and Eq.~(\ref{O(N):symmetry breaking:2}) implies,
\begin{equation}
m^2+\frac{\lambda}{N}\left(3\imath\Delta^d_{11}(x;x)_{\rm fin}+(N-1)\imath\Delta^d_{22}(x;x)_{\rm fin}\right)>0
\,
\label{O(N):symmetry unbroken}
\end{equation}
and analogous steps as above yield,
\begin{equation}
 \mu_1^2 = -\frac12 +\frac{\lambda_D}{2}\left(\frac{N\!-\!1}{\mu_g^2}+\frac{3}{\mu_1^2}\right)
 \,,\qquad
 \mu_g^2 = \mu_1^2 +\lambda_D\left(\frac{1}{\mu_g^2}-\frac{1}{\mu_1^2}\right)
\,.
\label{O(N):symmetry unbroken:2}
\end{equation}
Just as in the broken case, when $N=1$ we have,
\begin{equation}
  \mu_1^2 = -\frac14 - \sqrt{\frac{1}{16}+\frac{3\lambda_D}{2}}
\,,
\nonumber
\end{equation}
which agrees with Eq.~(\ref{mass gap equation:phi2=0:2}).
 In fact, it is quite easy to obtain the general solution of
 equations~(\ref{O(N):symmetry unbroken:2}). Indeed, observe that the second
 equation can be written as,
\begin{equation}
   (\mu_1^2-\mu_g^2)(\lambda_D+\mu_1^2\mu_g^2) = 0
\,.
\nonumber
\end{equation}
The positivity of $\lambda_D$, $\mu_1^2$ and $\mu_g^2$ immediately implies
that the only consistent solution is
\begin{equation}
  \mu_g^2 = \mu_1^2
\,.
\label{solution:unbroken:1}
\end{equation}
It is not surprising that in this case all particles must have the same mass
since the symmetry is unbroken. With this, the first equation
in~(\ref{O(N):symmetry unbroken:2}) is easily solved,
\begin{equation}
 \mu_1^2 = -\frac14-\sqrt{\frac{1}{16}+\frac{(N+2)\lambda_D}{2}}
\,,
\label{solution:unbroken:2}
\end{equation}
which can be also written as,
\begin{equation}
 M_1^2 = M_g^2 = \frac{m^2}{2}+\sqrt{\frac{m^4}{4}+\frac{(N+2)m_{\rm cr}^4}{6}}
\,,
 \label{solution:unbroken:3}
\end{equation}
where $m_{\rm cr}$ is given in~(\ref{critical mass}).
This generalizes the real field result~(\ref{mass gap equation:phi2=0:2})
to the $O(N)$ symmetric case. From Eq.~(\ref{solution:unbroken:3}) one can easily see
that $M_1^2 = M_g^2>0$, as it should be.
The solution~(\ref{solution:unbroken:3}) can be expanded
in powers of $\lambda$. When $m^2\gg \sqrt{N}m_{\rm cr}^2$
the expansion is analytic in $\lambda$,
$M_1^2=M_g^2\simeq m^2+(N+2)m_{\rm cr}^4/(3m^2)
      = m^2+(N+2)\lambda H^4/(8\pi^2 m^2)$,
while in the limit when $m\rightarrow 0$, the expansion is
non-analytic in $\lambda$,
$M_1^2=M_g^2\simeq \sqrt{[(N+2)/6]}\,m_{\rm cr}^2
     =\sqrt{[(N+2)\lambda]}\,H^2/(4\pi)$,
where in the latter equalities for simplicity we took $D=4$.

An important question is what does the jump in the order
parameter~(\ref{phase transition:O(N)}) imply for the nature of the transition.
We can answer this question by considering the
analogous two inflationary scenarios from the end of
section~\ref{A real scalar field}.
Just like in the real scalar field case,
in {\it Scenario A}, in which inflation
begins from a false vacuum state,
the transition is a crossover and proceeds
{\it via} temporary breaking of de Sitter symmetry,
until a de Sitter invariant state (over finite, but very large domains)
gets established. In {\it Scenario B},
in which inflation is preceded by a radiation era,
the transition can be of first order (if $|m^2|\gg H^2$),
bubbles of the broken phase nucleate and the field
tunnels to the broken phase minimum. During the transition
topological defects in general form (global cosmic strings when $N=2$,
global monopoles when $N=3$ and higher order defects such as
cosmic texture when $N\geq 3$).

 \section{Discussion}
 \label{Discussion}

 We have analysed the $O(N)$ symmetric scalar field model~(\ref{O(N) action})
in the mean field (Hartree) approximation~(\ref{O(N) effective potential})
on de Sitter space. We have shown that symmetry breaking can occur,
and that the would-be Goldstone bosons acquire a mass
(see figure~\ref{fig:masses}) due to the enhanced
infrared correlations in de Sitter space, and that the $O(N)$
symmetry gets completely broken by the ground state of the theory.
Next we have studied the strength
of the transition and shown that, depending
on the inflationary scenario assumed, the transition can proceed either
as a crossover or as a first order phase transition.
Curiously, the jump in the order parameter~(\ref{phase transition:O(N)})
exhibits a non-analytic dependence on
the parameters of the model, $\Delta\phi^2 \propto |m|N^{1/4}\lambda^{-3/4}$,
where $|m|$ and $\lambda$ denote the mass parameter and
the quartic self-coupling of
the model.

 While the mean field results are of their own interest,
it would be desirable to investigate whether (and how) the mean field results presented here
change when one includes higher loop corrections.
A first step in this direction is taken in
Ref.~\cite{Garbrecht:2011gu} where the local contribution to the self-mass
from the two loop (sun-set) diagram was estimated, and where it was found
that, in the massless limit, the mean field mass-squared gets reduced by
a factor $1/\sqrt{2}$.

 Second, it is instructive to compare our results with the (old) stochastic
theory results of Starobinsky and Yokoyama~\cite{Starobinsky:1994bd},
which is known to resum the leading $\log(a)$ corrections to infrared correlators
on de Sitter space, see {\it e.g.}~\cite{Prokopec:2007ak}.
From Eq.~(23) of Ref.~\cite{Starobinsky:1994bd} we read
(upon a rescaling, $\lambda\rightarrow\lambda/6$),
\begin{equation}
m^2_{\rm stoch}=\frac{\lambda}{2}\langle \phi^2\rangle
               =\frac{3}{2\pi}\frac{\Gamma(3/4)}{\Gamma(1/4)}\sqrt{\lambda}H^2
               \approx 0.1614\sqrt{\lambda}H^2
\,,
\label{stochastic mass}
\end{equation}
which is to be compared with Eqs.~(\ref{mass gap equation:phi2=0:2})
and~(\ref{critical mass}), which in the limit when $m^2\rightarrow 0$ and in $D=4$ yield
$m^2_{\rm MF}\rightarrow  \sqrt{3\lambda}H^2/(4\pi)$. This then implies,
\begin{equation}
\frac{m^2_{\rm stoch}}{m^2_{\rm MF}} = 2\sqrt{3}\frac{\Gamma(3/4)}{\Gamma(1/4)}\approx 1.17
\,.
\nonumber
\end{equation}
Even though the difference in the results is modest,
the question --  which result is correct? -- is, nevertheless, important.
In the derivation of the stochastic result~(\ref{stochastic mass}),
one assumes that the tree level
potential remains unchanged, {\it i.e.} that for the late time behaviour
the tree level potential should be used when stochastic theory is applied to inflation.
At the moment there is no fundamental understanding concerning whether
the tree level potential or some effective potential should be
used in stochastic formalism. We close this discussion by noting
that one can recover exactly the mean field result $(m_{\rm MF}^2)_{m\rightarrow 0}$
from stochastic formalism, provided one replaces the tree level potential
$V=(\lambda/4!)\phi^4$ by its Gaussian counterpart,
$V\rightarrow V_{\rm Gauss} = (\lambda/4)\langle(\phi)^2\rangle\phi^2$.
While this is suggestive, it does not ultimately tell us what is the correct answer.

Furthermore, it is useful to mention the well understood thermal case,
where also non-analytic behaviour in the coupling constant occurs
when a self-consistent Hartree approximation (daisy resummation) is employed
in the model considered in this paper.
Up to a logarithmic correction, in the symmetric case ($m^2>0$)
the resummed mass of a real scalar field of section~\ref{A real scalar field}
is of the form,
\begin{equation}
 m_{\rm MF}(T) = \sqrt{m^2+\frac{\lambda T^2}{24}
                         +\frac{\lambda^2 T^2}{(16\pi)^2}}
                   -\frac{\lambda T}{16\pi}
\,,
\nonumber
\end{equation}
which, when $m^2\rightarrow 0$ yields,
$m_{\rm MF}^2(T) \simeq \frac{\lambda T^2}{24}
     \Big(1 -\frac{\sqrt{3\lambda}}{2\sqrt{2}\pi} \Big)$.
The crucial difference with the de Sitter result~(\ref{solution:unbroken:3})
is that the thermal series for $m^2_{\rm MF}$ begins at $\sim\lambda T^2$,
and not at $\sim\sqrt{\lambda}H^2$ as it is the case in the de Sitter case.
This is because the infrared sector of de Sitter space is more infrared
divergent than the thermal infrared sector of bosonic field theory.

 Finally, after the original version of this work was completed,
an interesting paper by Boyanovsky appeared~\cite{Boyanovsky:2012nd},
which used somewhat different techniques and
confirmed the main results of this work, namely the
symmetry breaking in an $O(N)$ scalar model and the
 mass generation mechanism of (would-be) Goldstone bosons.

 \section*{Acknowledgments}

 I thank Bjorn Garbrecht, Louis Leblond, Albert Roura, Julien Serreau and
Richard Woodard for useful discussions. I am particularly grateful to
Julien for critical remarks on the manuscript and to Louis and Albert
for very useful discussions at the `Quantum Gravity: From UV to IR' 2011
workshop at CERN,
during which the basic idea of the paper was born.

\section*{Appendix: The de Sitter space propagator for massive and massless scalar fields}
\label{Appendix: The de Sitter space propagator for massive scalar fields}

 Here we review some of the basic properties of the scalar propagator on de Sitter background
in $D$ space time dimensions.
The time ordered (Feynman) propagator $\imath \Delta\equiv\imath \Delta^{++}$
obeys the equation,
\begin{equation}
 \sqrt{-g}\left[\Box_D-m^2\right]\imath \Delta^{++}(x;x^\prime) = \imath \delta^D(x-x^\prime)
\,,
\label{A:propagator}
\end{equation}
where $m$ is a mass and $\Box_x=(-g)^{-1/2}\partial_\mu g^{\mu\nu}\sqrt{-g}\partial_\nu$
is the scalar d'Alembertian in D spacetime dimensions. The de Sitter
invariance allows one to write the d'Alembertian in a de Sitter invariant form,
\begin{equation}
\Big[\bar y(4-\bar y)\frac{d^2}{d\bar y^2}
     + D(2-\bar y)\frac{d}{d\bar y}
  -\frac{m^2}{H^2}\Big]\imath\Delta(x;x^\prime)
 =\frac{\imath}{\sqrt{-g}H^2}\delta^D(x-x^\prime)
\,,
\label{A:propagator:2}
\end{equation}
where $\bar y = a(\eta)a(\eta^\prime)H^2[-(\eta-\eta^\prime)^2 +\|\vec x - \vec x^\prime\|^2]$
is related to the geodesic distance on de Sitter space $\ell(x;x^\prime)$ as,
$\bar y = 4\sin^2(H\ell/2)$. Here $a$ denotes the scale factor,
$\eta$ is conformal time
and $\vec x$ comoving coordinate.
The unique solutions for the relevant propagators of the Schwinger-Keldysh (or {\it in-in})
formalism can be written in terms of the Gauss' hypergeometric function ${}_2F_1$
as follows,
\begin{equation}
      \imath\Delta^{\alpha\beta}(x;x^\prime)=
      \frac{H^{D-2}}{(4\pi)^{D/2}}
      \frac{\Gamma(\frac{D-1}{2}+\nu_D)\Gamma(\frac{D-1}{2}-\nu_D)}{\Gamma(\frac{D}{2})}\\
 \times{}_2F_1\Big(\frac{D-1}{2}+\nu_D,\frac{D-1}{2}-\nu_D;\frac{D}{2};1-\frac{y^{\alpha\beta}}{4}\Big)
      \,,
\label{A:propagator:3}
\end{equation}
where
\begin{equation}
\nu_D^2=\Big(\frac{D-1}{2}\Big)^2-\frac{m^2}{H^2}
\,.
 \label{A:nuD}
\end{equation}
Here $m^2>0$ represents the (renormalised) field mass parameter, which
includes the renormalised mass, the mean field correction
(the finite part of $(\lambda/2)\imath \Delta(x;x)$)
and possibly also the term that originates
from a nonminimal coupling, $\Delta m^2 = \xi D(D-1)H^2$ in the lagrangian,
$\Delta {\cal L} = -\xi R \phi^2$, where $R = D(D-1) H^2$ is the Ricci scalar in de Sitter
space. The functions $y^{\alpha\beta}$ ($\alpha,\beta=\pm$) in Eq.~(\ref{A:propagator:3})
denote,
\begin{eqnarray}
y^{++} &=& a(\eta)a(\eta^\prime)H^2[-(|\eta-\eta^\prime|-\imath \epsilon)^2
                                +\|\vec x - \vec x^\prime\|^2]
\nonumber\\
y^{+-} &=& a(\eta)a(\eta^\prime)H^2[-(\eta-\eta^\prime+\imath \epsilon)^2
                                +\|\vec x - \vec x^\prime\|^2]
\nonumber\\
y^{-+} &=& a(\eta)a(\eta^\prime)H^2[-(\eta-\eta^\prime-\imath \epsilon)^2
                                +\|\vec x - \vec x^\prime\|^2]
\nonumber\\
y^{--} &=& a(\eta)a(\eta^\prime)H^2[-(|\eta-\eta^\prime|+\imath \epsilon)^2
                                +\|\vec x - \vec x^\prime\|^2]
\,,
\label{A:y:def}
\end{eqnarray}
with $\epsilon>0$ infinitesimal. All propagators in~(\ref{A:propagator:3})
have the same coincident limit,
\begin{equation}
\imath\Delta(x;x)=\imath\Delta^{\alpha\beta}(x;x)
  = \frac{H^{D-2}}{(4\pi)^{D/2}}
  \frac{\Gamma\left(\frac{D-1}{2}+\nu_D\right)\Gamma\left(\frac{D-1}{2}-\nu_D\right)}
       {\Gamma\left(\frac{1}{2}+\nu_D\right)\Gamma\left(\frac{1}{2}-\nu_D\right)}
        \Gamma\left(1-\frac{D}{2}\right)
\,.
\label{A:coincident limit}
\end{equation}
Due to the last $\Gamma$ function, this propagator exhibits a simple pole in even
dimensions, $D=2,4,6,..$, which reflects an ultraviolet (UV) logarithmic divergence.
Of course, the leading UV divergence of the coincident propagator
in de Sitter space is the same as that in Minkowski space, and it is of a degree $D-2$,
the subleading is of a degree $D-4$, {\it etc.}, the degree zero representing a logarithmic
divergence. As it is well known, dimensional regularisation is blind to power law divergences
(they are automatically subtracted by analytic extension), and exhibits only logarithmic
divergences. The effect of the propagator~(\ref{A:coincident limit}) can be
considered in the weak curvature (Minkowski) limit, when $m^2\gg H^2$
(in which case one recovers the Minkowski space result plus small corrections)
and in a strong curvature regime, in which $m^2\ll H^2$.
Ignoring the UV divergence in~(\ref{A:coincident limit}) one can naively expand
it in powers of $m^2/H^2$, and one obtains~\footnote{The ${\cal O}(m^2)$ term
in Eq.~(\ref{A:coincident limit:2}) has a divergent coefficient.
To make the renormalised gap equation consistent, the divergent part of
the ${\cal O}(m^2)$ term would have to be absorbed in the renormalised mass term $m^2$.},
\begin{eqnarray}
\imath\Delta(x;x)
  &=& \frac{H^{D-2}\Gamma\left(\frac{D\!-\!1}{2}\right)}{4\pi^{(D-1)/2}}
    \left[\psi\left(\frac{D}{2}\right)\!-\!\psi(D\!-\!1)
          \!-\!\psi\left(1\!-\!\frac{D}{2}\right)\!-\!\gamma_E+\frac{1}{D\!-\!1}
    \right]
\nonumber\\
  &+& \frac{\Gamma\left(\frac{D+1}{2}\right)}{2\pi^{(D+1)/2}}\frac{H^D}{m^2}
  + {\cal O}\Big(\frac{m^2}{H^2}\Big)
\,,
\label{A:coincident limit:2}
\end{eqnarray}
such that in $D=2,3,4$ the ${\cal O}(m^{-2})$ terms are $H^2/(4\pi m^2)$,
$H^3/(2\pi^2 m^2)$, and $3H^4/(8\pi^2 m^2)$,
respectively. In our analysis in the main text
we assume that both finite and infinite $m$-independent terms
in~(\ref{A:coincident limit:2})
are absorbed in the physical definition of the mass term.

 The de Sitter invariant limit will be attained after some time during inflation.
If the mass is very small ($m^2\ll H^2$), the propagator will at early times
grow logarithmically with the scale factor (linearly with cosmological time).
This can be seen by recalling
that in the infrared~\cite{Janssen:2008px}
the coincident propagator satisfies,
\begin{eqnarray}
 \imath\Delta(x;x) &=& \frac{H^{D-2}}{2^D\pi^{(D-3)/2}\Gamma\left(\frac{D-1}{2}\right)}
 \int_{k_0/(Ha)}^\infty dz z^{D-2}|H^{(1)}_{\nu_D}(z)|^2
\nonumber \\
 &=& \frac{H^{D-2}}{(4\pi)^{D/2}}
  \frac{\Gamma\left(\frac{D-1}{2}+\nu_D\right)\Gamma\left(\frac{D-1}{2}-\nu_D\right)}
       {\Gamma\left(\frac{1}{2}+\nu_D\right)\Gamma\left(\frac{1}{2}-\nu_D\right)}
        \Gamma\left(1-\frac{D}{2}\right)
 - \frac{H^{D-2}\Gamma\left(\nu\right)^2}{8\pi^{(D-3)/2}\Gamma\left(\frac{D-1}{2}\right)}
       \times    \frac{\left(\frac{k_0}{2Ha}\right)^{D-1-2\nu_D}}{D-1-2\nu_D}
\nonumber\\
&& +\,{\cal O}\left(k_0^{D+1-2\nu_D},k_0^{D-1},k_0^{D-1+2\nu_D}\right)
 \,,
\label{A:propagator:time dep}
\end{eqnarray}
where we took $a(t_0)=a_0=1$ and $k_0$ is an infrared (comoving) momentum cut-off.
Notice that at early times (and in the limit when $m\rightarrow 0$ and
$\nu_D\rightarrow (D-1)/2$)
the coincident propagator grows logarithmically with time as
(see also Refs.~\cite{Prokopec:2003tm,Prokopec:2004au})
\begin{eqnarray}
 \imath\Delta(x;x) &=&
\frac{H^{D-2}\Gamma\left(\frac{D\!-\!1}{2}\right)}{4\pi^{(D-1)/2}}
    \left[\psi\left(\frac{D}{2}\right)\!-\!\psi(D\!-\!1)
          \!-\!\psi\left(1\!-\!\frac{D}{2}\right)\!-\!\gamma_E+\frac{1}{D\!-\!1}
    \right]
\nonumber\\
   &+& \frac{H^{D-2}\Gamma\left(\frac{D-1}{2}\right)}{2\pi^{(D+1)/2}}
              \left[\ln(a)-\ln\left(\frac{k_0}{2H}\right)\right]
   +{\cal O}(m^2/H^2)
\,.
\label{A:propagator:time dep:2}
\end{eqnarray}
This is to be compared with the Onemli-Woodard coincident propagator for
a massless scalar field~\cite{Onemli:2004mb}
\footnote{Not surprisingly, the coincident propagator of a light scalar
field~(\ref{A:propagator:time dep}) and the Onemli-Woodard coincident massless scalar
propagator~(\ref{A:propagator:time dep:2}) possess identical late time logarithmically growing
terms $\propto\ln(a)$.
The time-independent parts do not agree, however. But this was to be expected,
since these constant pieces do not have an
independent physical meaning, as they can be absorbed in the mass counterterm
of the self-interacting scalar theory.
}:
\begin{equation}
 [\imath\Delta(x;x)]_{\rm OW} = \frac{H^{D-2}}{4\pi^{(D-1)/2}}
   \frac{\Gamma\left(\frac{D}{2}\right)\Gamma\left(1-\frac{D}{2}\right)}{\Gamma\left(\frac{3-D}{2}\right)}
   + \frac{H^{D-2}\Gamma\left(\frac{D-1}{2}\right)}{2\pi^{(D+1)/2}}\ln(a)
\,.
\label{A:propagator:time dep:3}
\end{equation}
The logarithmic growth saturates when the propagator reaches the de Sitter invariant
value~(\ref{A:coincident limit:2}), which characterizes
the time scale at which the propagator (and thereby the state) becomes de Sitter invariant.

\end{document}